\documentclass{ws-ijmpd}

\begin{document}

\markboth{S. M. Merkowitz, et. al.}
{Laser Ranging for Gravitational, Lunar, and Planetary Science}

%
\catchline{}{}{}{}{}
%

\title{Laser Ranging for Gravitational, Lunar, and Planetary Science}

\author{Stephen M. Merkowitz, Philip W. Dabney, Jeffrey C. Livas, Jan 
F. McGarry, Gregory A. Neumann, and Thomas W. Zagwodzki}

\address{NASA Goddard Space Flight Center, Greenbelt MD 20771, USA}

\maketitle

\begin{history}
\received{Day Month Year}
\revised{Day Month Year}
\comby{Managing Editor}
\end{history}

\begin{abstract}
More precise lunar and Martian ranging will enable unprecedented tests of
Einstein's theory of General Relativity and well as lunar and planetary
science.  NASA is currently planning several missions to return to the
Moon, and it is natural to consider if precision laser ranging
instruments should be included.  New advanced retroreflector arrays
at carefully chosen landing sites would have an immediate positive
impact on lunar and gravitational studies.  Laser transponders are
currently being developed that may offer an advantage over passive
ranging, and could be adapted for use on Mars and other distant objects.
Precision ranging capability can also be combined with optical communications for an extremely versatile instrument.  In this paper we discuss the science that can
be gained by improved lunar and Martian ranging along with several
technologies that can be used for this purpose.
\end{abstract}


\section{Introduction}

Over the past 35 years, lunar laser ranging (LLR) from a variety of
observatories to retroreflector arrays placed on the lunar surface by
the Apollo astronauts and the Soviet Luna missions have dramatically
increased our understanding of gravitational physics along with Earth
and Moon geophysics, geodesy, and dynamics.  During the past few
years, only the McDonald Observatory (MLRS) in Texas and the
Observatoire de la C™te d'Azur (OCA) in France have routinely made
lunar range measurements.  A new instrument, APOLLO, at the Apache
Point facility in New Mexico is expected to become operational within
the next year with somewhat increased precision over previous
measurements.\cite{Murphy_Ranging_2000}

\begin{figure}[htbp]
\begin{center}
\includegraphics[height=2in]{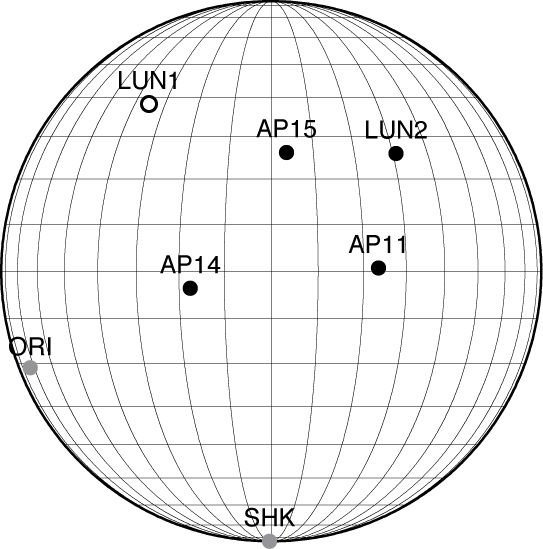}
\caption{Location of the lunar retroreflector arrays.  The three
Apollo arrays are labeled AP and the two Luna arrays are labeled
LUN. ORI and SHK show the potential locations of two additional
sites that would aid in strengthening the geometric coverage.}
\label{fig:sites}
\end{center}
\end{figure}

Setting up retroreflectors were a key part of the Apollo missions so
it is natural to ask if future lunar missions should include them as
well.  The Apollo retroreflectors are still being used today, and the
35 years of ranging data has been invaluable for scientific as well as
other studies such as orbital dynamics.  However, the available
retroreflectors all lie within 26 degrees latitude of the equator, and
the most useful ones within 24 degrees longitude of the sub-earth
meridian as shown in Fig.~\ref{fig:sites}.  This clustering weakens their
geometrical strength.  New retroreflectors placed at locations other
than the Apollo sites would enable more detailed studies, particularly
those that rely on the measurement of lunar librations.  In addition,
more advanced retroreflectors are now available that will reduce some
of the systematic errors associated with using the Apollo arrays.

In this paper we discuss the possibility of putting advanced
retroreflectors at new locations on the lunar surface.  In addition,
we discuss several active lunar laser ranging instruments that have
the potential for greater precision and can be adapted for use on
Mars.  These additional options include laser transponders and laser
communication terminals.

\section{Gravitational Science From Lunar Ranging}

Gravity is the force that holds the universe together, yet a theory
that unifies it with other areas of physics still eludes us.  Testing
the very foundation of gravitational theories, like Einstein's theory
of General Relativity, is critical in understanding the nature of
gravity and how it relates to the rest of the physical world.

The Equivalence Principle, which states the equality of gravitational
and inertial mass, is central to the theory of General Relativity.
However, nearly all alternative theories of gravity predict a
violation of the Equivalence Principle.  Probing the validity of the
Equivalence Principle is often considered the most powerful way to
search for new physics beyond the standard model.\cite{Damour_CQG_1996}  A violation of
the Equivalence Principle would cause the Earth and Moon to fall at
different rates toward the Sun resulting in a polarization of the
lunar orbit.  This polarization shows up in LLR as a displacement
along the Earth-Sun line with a 29.53 d synodic period.  The current
limit on the Equivalence Principle is given by LLR: 
$\Delta((M_{G}/M_{I})_{EP} = (-1.0\pm1.4)\times10^{-13}$.\cite{Williams_PRL_2004}

General Relativity predicts that a gyroscope moving through curved
spacetime will precess with respect to the rest frame.  This is
referred to as geodetic or de Sitter precession.  The Earth-Moon
system behaves as a gyroscope with a predicted geodetic precession of
19.2 msec/year.  This is observed using LLR by measuring the lunar
perigee precession.  The current limit on the deviation of the
geodetic procession is: $K_{gp}=(-1.9\pm6.4)\times10^{-3}$.\cite{Williams_PRL_2004}  This
measurement can also be used to set a limit on a possible cosmological
constant: $\Gamma < 10^{-26} \rm{km}^{-2}$.\cite{Sereno_PRD_2006}

It is also useful to look at violations of General Relativity in the
context of metric theories of gravity.  Post-Newtonian
Parameterization (PPN) provides a convenient way to describe simple
deviations from General Relativity.  The PPN parameters are usually
denoted as $\gamma$ and $\beta$; $\gamma$ indicates how much spacetime
curvature is produced per unit mass, while $\beta$ indicates how
nonlinear gravity is (self-interaction).  $\gamma$ and $\beta$ are
identically one in General Relativity.  Limits on $\gamma$ can be set from
geodetic procession measurements, but the best limits come from
measurements of the gravitational time delay of light, often referred
to as the Shapiro effect.  Ranging measurements to the Cassini
spacecraft set the current limit on $\gamma$: $(\gamma-1) =
(2.1\pm2.3)\times10^{-5}$,\cite{Bertotti_Nature_2003} which combined with LLR data provides
the best limit on $\beta$: $(\beta-1) = (1.2\pm1.1)\times10^{-4}$.\cite{Williams_PRL_2004}

The strength of gravity is given by Newton's gravitational constant G.
Some scalar-tensor theories of gravity predict some level of time
variation in G. This will lead to an evolving scale of the solar
system and a change in the mass of compact bodies due to a variable
gravitational binding energy.  This variation will also show up on
larger scales, such as changes in the angular power spectrum of the
cosmic microwave background.\cite{Uzan_RMP_2003}  The current limit on the time
variation of G is given by LLR: 
$\dot{G}/G =(4\pm9)\times10^{-13}$/year.\cite{Williams_PRL_2004}

The above effects are the leading gravitational limits that have been
set by LLR, but many more effects can be studied using LLR data at
various levels.  These include gravitomagnetism (frame-dragging),
$1/r^{2}$ force law, and even tests of Newton's third law.\cite{Nordtvedt_CQG_2001}

\section{Lunar Science From Lunar Ranging}

Several areas of lunar science are aided by LLR. First, the
orientation of the Moon can be used for geodetic mapping.  The current
IAU rotation model, with respect to which images and altimetry are
registered, has errors at the level of several hundred meters.  A more
precise model, DE403,\cite{Standish_JPL_1995} is being considered that is based on LLR
and dynamical integration, but will require updating since it uses
data only through 1994.  Errors in this model are believed to be
several meters.  Further tracking will quantify the reliability of
this and future models for lunar exploration.

Second, LLR helps provide the ephemeris of the Moon and solar system
bodies.  The position of the lunar center-of-mass is perturbed by
planetary bodies, particularly Venus and Jupiter, at the level of
100's of meters to more than 1 km.  LLR is an essential constraint on
the development of planetary ephemerides and navigation of spacecraft.

LLR can also be used to study the internal structure of the Moon, such
as the possible detection of a solid inner core.  The second-degree
tidal lunar Love numbers are detected by LLR, as well as their phase
shifts.  From these measurements, a fluid core of ~20\% the Moon's
radius is suggested.  A lunar tidal dissipation of $Q = 30\pm 4$ has
been reported to have a weak dependence on tidal frequency.  Evidence
for the oblateness of the lunar fluid-core/solid-mantle boundary may
be reflected in a century-scale polar wobble frequency.  The lunar
vertical and horizontal elastic tidal displacement Love numbers $h_{2}$
and $l_{2}$ are known to no better than 25\% of their values, and the
lunar dissipation factor $Q$ and the gravitational potential tidal
Love number $k_{2}$ no better than 11\%.  These values have been
inverted jointly for structure and density of the core,\cite{Khan_JGR_2004,Khan_GRL_2005}
implying a semi-liquid core and regions of partial melt in the lunar
mantle, but such inversions require stochastic sampling and yield
probabilistic outcomes.

\section{Rationale for Additional Lunar Ranging Sites}

While a single Earth ranging station may in principle range to any of the 
four usable retroreflectors on the near side from any longitude during
the course of an observing day, these observations are nearly the same
in latitude with respect to the Earth-Moon line, weakening the
geometric strength of the observations.  Additional observatories
improve the situation somewhat, but of stations capable of ranging to
the Moon, only Mt.  Stromlo in Australia is not situated at similar
northern latitudes.  The frequency and quality of observations varies
greatly with the facility and power of the laser employed.  Moreover,
the reflector cross sections differ substantially.  The largest
reflector, Apollo 15, has 300 cubes and returns only a few photons per
minute to MLRS. The other reflectors have 100 cubes or less, and
proportionately smaller rates.  Stations and reflectors are unevenly represented, so that in
recent years, most ranging has occurred between one ground station and
one reflector.  Over the past six years, 85\% of LLR data has been taken from MLRS and 15\% from OCA.  81\% of these were from the Apollo 15 reflector, 10\% from Apollo 11, 8\% from Apollo 14, and about 1\% from Luna 2.\cite{Shelus_Personal}  The solar noise background and thermal distortion makes ranging to some reflectors possible only around the quarter-moon
phase.  The APOLLO instrument should be capable of ranging during all
lunar phases.

The first LLR measurements had a precision of about 20 cm.  Over the
past 35 years, the precision has increased only by a factor of 10.
The new APOLLO instrument has the potential to gain another factor of
10, achieving mm level precision, but this capability has not yet been
demonstrated.\cite{Williams_IJMPD_2004}  Poor detection rates are a major limiting factor in
past LLR. Not every laser pulse sent to the Moon results in a detected
return photon, leading to poor measurement statistics.  MLRS typically
collects less than 100 photons per range measurement with a scatter of
about 2 cm.  The large collecting area of the Apache Point telescope
and the efficient avalanche photodiode arrays used in the APOLLO
instrument should result in thousands of detections (even multiple
detections per pulse) leading to a potential statistical uncertainty
of about 1 mm.  Going beyond this level of precision will likely
require new lunar retroreflectors or laser transponders that are more
thermally stable and are designed to reduce the error associated with
the changing orientation of the array with respect to the Earth due to
lunar librations.

Several tests of General Relativity and aspects of our understanding
of the lunar interior are currently limited by present LLR
capabilities.  Simply increasing the precision of the LLR measurement,
either through ground station improvement or through the use of laser
transponders, will translate into improvements in these areas.

Additional ranging sites will also help improve the science gained
through LLR. The structure and composition of the interior require
dynamic measurements of the lunar librations, while tests of General
Relativity require the position of the lunar center of mass.  In all,
six degrees of freedom are required to constrain the geometry of the
Earth-Moon system (in addition to Earth orientation).  A single
ranging station and reflector is insufficient to accurately determine
all six, even given the rotation of the Earth with respect to the
Moon.

To illustrate the importance of adding high-cross-section reflectors
(or transponders) near the lunar limb, we performed an error analysis
based on the locations of two observing stations that are currently
operating and the frequency with which normal points have been
generated over the last 10 years.  While data quality has improved
over the years, it has reached a plateau for the last 15 years or so.
The presently operating stations are comparable in quality, and we
assume an average of 2.5 cm for all observations.  The normal point
accumulation is heavily weighted toward Apollo 15, and a
negligible number of returns are obtained from Lunakhod 2.  We
anticipate that ranging to a reflector with 4x higher cross section
than Apollo 15 would approach 1 cm quality, simply by the increased return rate.

The model assumes a fixed Earth-Moon geometry to calculate the
sensitivity of the position determination jointly with lunar rotation
along three axes parallel to Earth's X-Y-Z coordinate frame at a
moment in time when the Moon lies directly along the positive X axis.
The Z axis points North and Y completes the right-hand system.
Partial derivatives of range are calculated with respect to
perturbations in position and orientation, where orientation is scaled
from radians to meters by an equatorial radius of 1738 km.  The
analysis makes no prior assumptions regarding the dynamical state of
the Moon.

The normal equations are weighted by the frequency of observations at
each pair of ground stations and reflectors over the last ten years.  We then replace some of the
observations with ranges to one or more new reflectors.  The results
are given in Table.~\ref{tab:precision}.

\begin{table}[h]
\tbl{Additional ranging sites and stations increase the precision
of the normal points for the same measurement precision due to better
geometrical coverage.  The precision in meters on the six degrees of
freedom of a typical normal point is shown for several possible
ranging scenarios.}
{\begin{tabular}{@{}|l|c|c|c|c|c|c|l|@{}} \hline
X & Y & Z & RotX & RotY & RotZ & \\ \hline
0.265 & 6.271 & 23.294 & 15.958 & 0.179 & 0.225 & MLRS and OCA \\ \hline
0.263 & 3.077 & 23.305 & 7.611 & 0.174 & 0.140 & 25\% of observations to ORI \\ \hline
0.259 & 2.840 & 23.271 & 4.692 & 0.114 & 0.198 & 25\% of  observations to SHK \\ \hline
0.259 & 2.969 & 23.291 & 4.850 & 0.116 & 0.086 & both ORI and SHK \\ \hline
0.030 & 2.501 & 2.902 & 4.244 & 0.050 & 0.078 & both ORI and SHK with 25\% \\
&&&&&&additional observations from \\
&&&&&&Mt. Stromlo \\ \hline
\end{tabular} \label{tab:precision}}
\end{table}

The addition of one or more reflectors would improve the geometrical
precision of a normal point by a factor of 1.5 to nearly 4 at the same
level of ranging precision.  Such improvements directly scale to
improvements in measurement of ephemeris and physical librations.  The
uncertainty in Moon-Earth distance (X) is highly correlated with
uncertainty in position relative to the Ecliptic (Z) when all stations
lie at similar latitudes.  An advanced reflector with high cross
section would enable southern hemisphere ground stations such as Mt.
Stromlo to make more frequent and precise observations.  The geometric
sensitivity to position is dramatically improved by incorporating such
a ground station, as shown in the last row of Table.~\ref{tab:precision}.

\section{Retroreflectors}

Five retroreflector arrays were placed on the Moon in the period 1969
- 1973.  Three were placed by US astronauts during the Apollo missions
(11, 14, and 15), and two were sent on Russian Lunokhod landers.  The
Apollo 11 and 14 arrays consist of 100 fused silica ``circular opening''
cubes (diameter 3.8 cm each) with a total estimated lidar cross
section of ~0.5 billion square meters.  Apollo 15 has 300 of these
cubes and therefore about 3 times the lidar cross section and is the lunar array with the highest response.  Because
the velocity aberration at the Moon is small, the cube's reflective
face angles were not intentionally spoiled (deviate from 90 degrees).

The two Lunokhod arrays consist of 14 triangular shaped cubes, each
side 11cm.  Shortly after landing, the Lunokhod 1 array ceased to be a
viable target - no ground stations have since been able to get returns
from it.  It is also very difficult to get returns from Lunokhod 2
during the day.  The larger size of the Lunokhod cubes makes them less
thermally stable which dramatically reduces the optical performance
when sunlit.

Since 1969, multiple stations successfully ranged to the lunar
retroreflectors.  Some of these stations are listed in
Table.~\ref{tab:ground} along with their system characteristics.
However, there have only been two stations continuously ranging to the
Moon since the early 1970s: OCA in
Grasse, France, and MLRS in Texas.
The vast majority of their lunar data comes from the array with the
highest lidar cross section - Apollo 15.

The difficulty in getting LLR data is due to the distance to the Moon
coupled with the $1/r^4$ losses in the signal, and the technology
available at the ground stations.  MLRS achieves an expected return
rate from Apollo 15 of about one return per minute.  Increasing the
lidar cross section of the lunar arrays by a factor of 10 would
correspond to a factor of 10 increase in the return data rate.  This
can be achieved by making arrays with 10 times more cubes than Apollo
15 or by changing the design of the cubes.  One possibility is
increasing the cube size.  The lidar cross section of a cube with a
diameter twice that of an Apollo cube would be 16 times larger.
However, simply making solid cubes larger increases their weight by
the ratio of the diameter cubed.  Additional size also adds to thermal
distortions and decreases the cube's divergence: a very narrow
divergence will cause the return spot to completely miss the station
due to velocity aberration.  Spoiling can compensate for the velocity
aberration but reduces the effective lidar cross section.  Changing
the design of the cubes, such as making them hollow, may be a better
alternative.  For example, 300 unspoiled 5 cm beryllium hollow cubes
would have a total mass less than that of Apollo 15 but would have ~3x
higher lidar cross section.

\begin{figure}[htbp]
\begin{center}
	\includegraphics[height=2in]{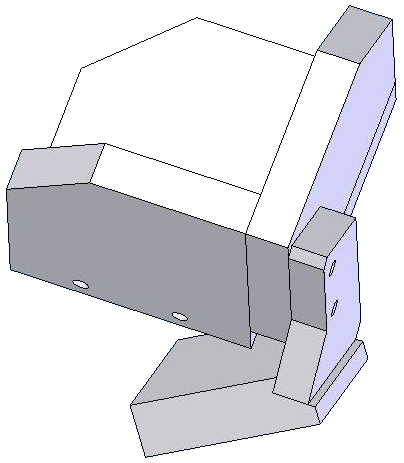}
	\caption{Hollow retroreflectors can potentially be used to build
	large cross-section lightweight arrays.}
	\label{fig:cube}
\end{center}
\end{figure}  

An option being investigated at Goddard is to replace solid glass
cubes with hollow cubes which weigh much less than their solid
counterparts.  Thermal distortions are less, especially in hollow
cubes made of beryllium, so the cubes can be made larger without
sacrificing optical performance.  Hollow cubes (built by PLX) flew on the Japanese ADEOS satellite and on the Air Force Relay
Mirror Experiment, but are generally not used on satellites for laser
ranging.  This is due in part to the lack of optical performance test
data on these cubes under expected thermal conditions, but also
because of early investigations which showed that hollow cubes were
unstable at high temperatures.  Advances in adhesives and other
techniques for bonding hollow cubes make it worthwhile to
reinvestigate them.  Testing that was done for Goddard by ProSystems
showed that hollow cubes (with faces attached via a method that is
being patented by ProSystems) can survive thermal cycles from room
temperature to 150 degrees Celsius.  Testing has not yet been done at
cold temperatures.  Preliminary mechanical analysis 
indicate that the optical performance of hollow Beryllium cubes would
be more than sufficient for laser ranging.

\section{Satellite Laser Ranging Stations}

Satellite Laser Ranging began in 1964 at NASA's Goddard Space Flight
Center.  Since then it has grown into a global effort, represented by
the International Laser Ranging Service (ILRS)\cite{Pearlman_ASR_2002} of which NASA is a
participant.  The ILRS includes ranging to Earth orbiting artificial
satellites, ranging to the lunar reflectors, and is actively working
toward supporting asynchronous planetary transponder ranging.

The ILRS lunar retroreflector capable stations have event timers with
precisions of better than 50 picoseconds, and can tie their clocks to
UTC to better than 100 nanoseconds.  Most have arc-second tracking
capabilities and large aperture telescopes ($> 1$ meter).  Their lasers
have very narrow pulse widths ($< 200$ psec) and most have high energy
per pulse ($> 50$ mJ).  All have the ability to narrow their transmit
beam divergence to less than 50 µrad.  The detectors have a relatively
high quantum efficiency ($> 15$\%).  All current LLR systems range at
532nm.

Clearly there is more than one way to increase the laser
return rate from the Moon.  One is to deploy higher response
retroreflector arrays or transponders on the Moon.  Another is to
increase the capability of the ground stations.  A third is to add
more lunar capable ground stations.  A combination of all these
options would have the biggest impact.

The recent development of the Apache Point system, APOLLO,\cite{Murphy_Ranging_2000} shows
what a significant effect improving the ground station can make.
Apache Point can theoretically achieve a thousand returns per minute
from Apollo 15 versus the few per minute return rate from MLRS (see
Table~\ref{tab:ground}).  Apache Point does this by using a very
large aperture telescope, a somewhat higher laser output energy and
fire rate, and a judicious geographical location (where the
astronomical seeing is very good).  Other areas that could also
improve ground station performance are higher quantum efficiency
single photon detectors ( $> 30$\% QE at 532nm), higher repetition rate
lasers (kilohertz versus tens of hertz), and the use of adaptive
optics to maintain tight beam control.

\begin{table}[h]
\tbl{Lunar retroreflector capable laser ranging stations and their
expected return rate from the Apollo 15 lunar array.  Link
calculations use 1 billion meters squared for Apollo 15's cross
section, a mount elevation of 30 degrees, and a standard clear
atmosphere (transmission = 0.7 at zenith) for all but Apache Point
where transmission = 0.85 at zenith.  The laser divergence was taken
to be 40 µrad for MLRS and 20 arcsec for the other systems.  The
detector quantum efficiency was assumed to be 30\% for all systems.}
{\begin{tabular}{@{}|lccccc|@{}} \hline
System & Telescope    & Pulse energy & Laser fire & System         & Apollo 15 \\
           & aperture       & exiting         & rate (Hz)  & transmission & photoelectrons/min \\
           & (m)             & system (J)    &                &                    & link calculation \\
MLRS & 0.76 & 60 & 10 & 0.5 & 4 \\
OCA (France) & 1.54 & 60 & 10 & 0.22 & 20 \\
Matera (Italy) & 1.5 & 22 & 10 & 0.87 & 60 \\
Apache Point & 3.5 & 115 & 20 & 0.25 & 1728 \\ \hline
\end{tabular} \label{tab:ground}}
\end{table}

Higher cross section lunar retroreflectors may make it possible to use
NASA's next generation of satellite laser ranging stations (SLR2000)
for LLR. The prototype SLR2000 system is currently capable of single
photon asynchronous laser transponder ranging, and will participate in
both a 2-way asynchronous transponder experiment in 2007 and the 1-way
laser ranging to the Lunar Reconnaissance Orbiter (LRO) in 2008-2009.
Approximately ten SLR2000 stations are expected to be built and
deployed around the world in the coming decade.  Adding ten lunar laser
ranging stations to the existing few would dramatically increase the
volume of data as well as giving the data a wide geographical
distribution.  The global distribution of the new SLR2000 stations
would be very beneficial to data collection from an asynchronous
transponder on the Moon.

\section{Laser Transponder}

Laser transponders are currently being developed for satellite
ranging, but they can also be deployed on the lunar surface.
Transponders are active devices that detect an incoming signal,
respond with a known or predictable response signal, and are used to
either determine the existence of the device or positioning
parameters, such as range and/or time.  For extraterrestrial
applications, a wide range of electromagnetic radiation, such as radio
frequency (RF), are used for this signal.  To date, most spacecraft
are tracked using RF signals, particularly in the S and X bands of the
spectrum.  NASA and several other organizations routinely track Earth
orbiting satellites using optical satellite laser ranging (SLR).
Laser transponders have approximately a $1/r^{2}$ link advantage over
direct ranging loss of $1/r^{4}$, essentially because the signal is
propagating in only one direction before being regenerated.  In fact,
it is generally considered that ranging beyond lunar distances is not
practical using direct optical ranging to cube-corner reflectors.
Laser transponders are in general more energy and mass efficient than
RF transponders since they can work at single photon detection levels
with much smaller apertures and beam divergences.  A smaller beam
divergence has the added benefit that there is less chance of
interference with other missions, as well as making the link more
secure should that be necessary.  With the development and inclusion
of laser communications for spaceflight missions, it is logical to
include an optical transponder that uses the same opto-mechanical
infrastructure such that it has minimal impact on the mission
resources.

The simplest conceptual transponder is the synchronous or echo
transponder.  An echo transponder works by sending back a timing
signal with a fixed delay from the receipt of the base-station signal.
This device has the potential for the lowest complexity and autonomous
operations with no RF or laser based communications channel.  To
enable this approach, an echo pulse must be created with a fixed
offset delay that has less than 500 ps jitter from the arrival of the
Earth station signal.  This is very challenging given the current
state-of-the-art in space-qualifiable lasers.  Furthermore, several
rugged and simple laser types would be excluded as candidates due to
the lack of precision control of the pulse generation.  The
synchronous/echo transponder has a total link probability that is the
joint probability of each direction's link probability (approximately
the product of each).

\begin{figure}[htbp]
\begin{center}
	\includegraphics{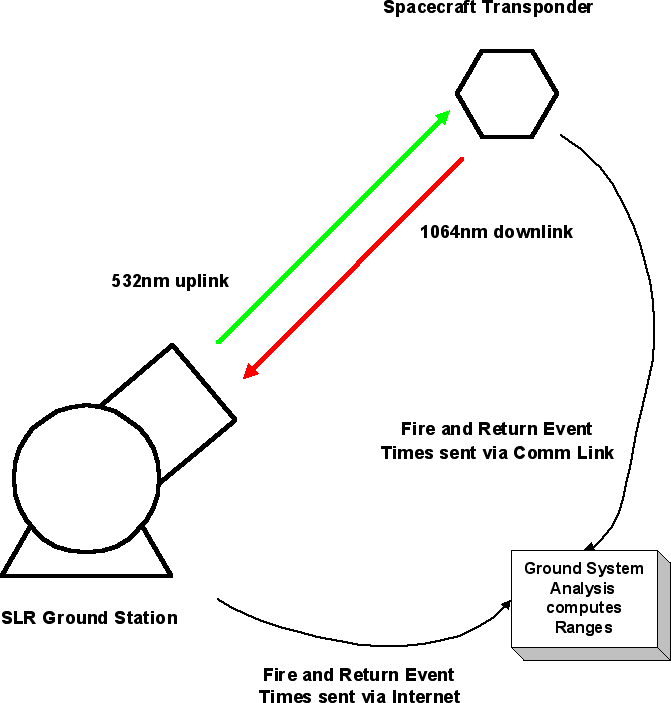}
	\caption{In an asynchronous laser transponder system the ground
	and remote stations fire independent of each other recording the
	pulse transmission and detection times.  The data from the two
	sites are then combined to calculate the range.}
	\label{fig:alt}
\end{center}  
\end{figure}

Asynchronous Laser Transponders (ALT) have been shown analytically\cite{Degnan_JGD_2002} and experimentally\cite{Smith_Science_2006} 
to provide the highest link probability
since the total link is the root-sum-square of each one-way link
probability.  Furthermore, they allow the use of free-running lasers
on the spacecraft that operate at their most efficient repetition
rates, are simpler, and potentially more reliable.  Fig.~\ref{fig:alt}
shows a conceptual asynchronous laser transponder using an existing
NASA SLR ground station that is already precisely located and
calibrated in the solar reference frame, and a spacecraft transponder
that receives green photons (532nm) and transmits near-infrared (NIR)
photons (1064 nm).  This diagram shows the spacecraft event times
being down linked on the RF (S-band) channel but this could be done on
the laser communication channel if one exists.  This dual wavelength
approach is being explored for reasons of technical advantage at the
ground station, but may also be used to help remove atmospheric
effects from the range data (due to its wavelength dependent index of
refraction).  Using the same wavelength for each direction is also
possible.  Expressions for recovering the range parameters from an
asynchronous measurement can be found in reference \refcite{Degnan_JGD_2002} and in the
parameter retrieval programs developed by Gregory Neumann for the
Earth-MLA asynchronous transponder experiments.\cite{Smith_Science_2006,Neumann_Ranging_2006}
 
An ALT will likely have systematic errors that will limit its
long-term accuracy.  A retroreflector array located near the ALT's
lunar site should allow the study and calibration of the ALT's
systematic errors.  Performing this experiment on the Moon will be
particularly important should this technology be adapted for Mars or
other bodies where retroreflectors cannot be used.

Recently, two interplanetary laser transponder experiments were
successfully demonstrated from the NASA Goddard Geophysical and
Astronomical Observatory (GGAO) SLR facility.  The first utilized the
non-optimized Mercury Laser Altimeter (MLA) on the Messenger
spacecraft and the second utilized the Mars Orbiting Laser Altimeter
(MOLA) on the Mars Orbiter spacecraft.  The Earth-MOLA experiment was
a one-way link that set a new distance record of 80 M-km for detected
signal photons.  The Earth-MLA experiment was a two way experiment
that most closely resembles the proposed asynchronous laser
transponder concept.  This experiment demonstrated the retrieval of
the clock offset, frequency drift, and range of 24 M-km using a small
number of detected two-way events.  These experiments have proven the
concept of being able to point both transceivers, detect the photons,
and retrieve useful parameters at low-link margins.

The Lunar Reconnaissance Orbiter (LRO) mission includes a GSFC
developed laser ranger that will provide a one-way ranging capability.
In this case the clock is assumed to be stable enough over one Lunar
orbit.  The result is a range profile that is extremely precise but
far less accurate than what a two-way asynchronous transponder would
provide with its full clock solution.

An ALT conceptual design was developed as part of the LRO laser ranger
trade study.  Link analyses performed on this design showed that it is
possible to make more than 500 two-way range measurements per-second
using a 20 mm aperture and a 10 micro-joule/pulse, 10-kHz laser at the
Moon and the existing eye-safe SLR2000 telescope located at the GGAO.
The ALT was not selected due to the need for a very high readiness
design for LRO, but the analysis did show its feasibility.  It was
also shown that many of the international SLR systems could
participate with nominal receiver and software upgrades thereby
increasing the ranging coverage.  The ALT increases the
tracking/ranging availability of spacecraft since the link margins are
higher than for direct ranging to reflector arrays.

\section{Communication Terminal}

A communications terminal conceptually represents the most capable
kind of transponder ranging system and is at the other end of the
spectrum in terms of complexity from the echo transponder.  In
general, a communications link of some kind is necessary to operate
and recover data from a spacecraft or remote site.  There are several
potential benefits if the communications link can be made part of the
ranging system, including savings in weight, cost, and complexity over
implementations that use separate systems for each requirement.

As with other types of transponder systems, the active terminals for a
full-duplex communications system mean that the loss budget for the
ranging/communications link scales as $1/r^{2}$ instead of $1/r^{4}$,
which is a substantial advantage.  The communications link need not be
symmetric in terms of data rate to achieve this benefit.  Very often
the uplink is relatively low bandwidth for command and control, and
the downlink is at a higher data rate for dumping data.  The ground
antenna can be made substantially larger than the remote antenna both
to make the receiver more sensitive and as a way to reduce the mass
and the pointing and tracking requirements for the spacecraft, since a
smaller antenna has a larger beam.

Forward Error Correction (FEC) is a technique that can improve the
link budget and hence the range of the system.  FEC is a signal
processing technique that adds additional bits to the communications
data stream through an algorithm that generates enough redundancy to
allow these bits to be used to detect errors.  There are a wide
variety of possible FEC algorithms that can be used, but it is
possible to get link budget gains of the order of 8 dB at the cost of
7\% overhead on the data rate even at data rates of several Gbits/sec.
Gains of the order of 10 dB and higher are available at lower data
rates, but at the cost of higher overhead.  Generally the FEC
algorithm may be optimized for the noise properties of the link if
they are known.

A synchronous communications terminal must maintain a precise clock to
be able to successfully recover the data.  A remote terminal will
recover the clock from the incoming data stream and phase-lock a local
oscillator.  All modern wide area terrestrial communications networks
use synchronous techniques, so the techniques and electronics are well
known and generally available.  The advantage of having a stable
reference clock that is synchronized to the ground terminal is that
long times (and therefore long distances) may be measured with a
precision comparable to that of the clock simply by counting bits in
the data stream and carefully measuring the residual timing offset at
the ground station.  A maximal-length pseudorandom code can be used to
generate a pattern with very simple cross-correlation properties that
may be used to unambiguously determine the range, and the synchronous
nature of the signal plus any framing or FEC structure imposed on the
data stream mean that even long times may be measured with the same
precision as the clock.

For optical communications terminals, it is almost as cost-effective
to run at a high data rate as it is at a low data rate.  The data rate
might then reasonably be chosen for the timing precision instead of
for the data downlink requirements.  For example, a 10 Gbps data rate
has a clock period of 100 picoseconds, which translates to a
sub-millimeter distance precision with some modest averaging - just
based on the clock.  Specialized modulation formats such as
phase-shift keying offer the possibility of optical phase-locking in
addition to electrical phase-locking, which may allow further
increases in precision.

Spacecraft or satellites may be used as repeaters or amplifiers much
as they are in terrestrial telecom applications, further extending the
reach.  Multiple communications terminals distributed around an
object, such as a planet, offer the ability to measure more
complicated motion than just a range and a change in range.  A high
data rate terminal might also be used as part of a communications
network in space.  In addition to serving as a fixed point for high
precision ranging, it could also provide various communications
functions such as switching and routing.

\section{Conclusions}

LLR has made great advances in the past 35 years.  However, the amount
of light returned by the current retroreflectors is so little that
only the largest ranging stations can be used for this purpose; poor
detection statistics remains the leading source of error.  Thermal and
orientation effects will ultimately limit range measurement to the
Apollo retroreflectors.  Measurements of the lunar librations are also
limited by the poor geometric arrangement of the visible
retroreflectors.

More precise range measurements to retroreflectors placed at sites far
from the existing arrays will greatly improve the gravitational and
lunar science discussed above.  A number of improvements (such as
higher cross section) can be made to the retroreflector designs to
realize these gains.  This natural extension to the Apollo instruments
is likely to produce a solid incremental improvement to these
scientific studies for many years to come.

To make a much larger leap in ranging accuracy, a laser transponder or
communication terminal will most likely be required.  The robust link
margins will enable the use of much smaller ground stations, which
would provide for more complete time and geometric coverage as more
ranging stations could be used.  An active system will also not be
susceptible to the libration induced orientation errors.

An active laser ranging system can be considered a pathfinder for a
Mars instrument, as it is likely to be the only way to exceed the
meter level accuracy of current ranging data to Mars.  Laser ranging to Mars can be used to measure the gravitational time
delay as Mars passes behind the Sun relative to the Earth.  With 1 cm
precision ranging, the PPN parameter $\gamma$ can be measured to about
$10^{-6}$, ten times better than the Cassini result.\cite{Turyshev_Exploration_2004}  The Strong
Equivalence Principle polarization effect is about 100 times larger
for Earth-Mars orbits than for the lunar orbit.  With 1 cm precision
ranging, the Nordtvedt parameter, $\eta = 4\beta-\gamma-3$, can be
measured to between $6\times10^{-6}$ and $2\times10^{-6}$ for
observations ranging between one and ten years.\cite{Anderson_APJ_1996}  Combined with
the time delay measurements this leads to a measurement of PPN
parameter $\beta$ to the $10^{-6}$ level.  Mars ranging can also be
used in combination with lunar ranging to get more accurate limits on
the time variation of the gravitational constant.

The ephemeris of Mars itself is known to meters in plane, but hundreds
of meters out-of-plane.\cite{Konopliv_Icarus_2006}  Laser ranging would get an order of
magnitude better estimate, significant for interplanetary navigation.
Better measurements of Mars' rotational dynamics could provide
estimates of the core size.\cite{Folkner_Science_1997}  The elastic tidal Love number
is predicted to be less than 10 cm, within reach of laser ranging.
There is also an unexplained low value of $Q$, inferred from the secular
decay of Phobos' orbit, that is a constraint to the present thermal
state of the Mars interior \cite{Bills_JGR_2005}.  Laser ranging to Phobos would help
solve this mystery.

\end{document}